\documentclass{optica-article}

\journal{opticajournal} 

\articletype{Research Article}
\usepackage{array}
\usepackage{lineno}
\usepackage[normalem]{ulem}
 
\begin{document}

\title{Long wave infrared detection using spintronic poisson bolometer arrays}

\author{Utkarsh Singh,\authormark{1} Leif Bauer,\authormark{1}, Angshuman Deka,\authormark{1} Mohamed Mousa,\authormark{1} Daien He,\authormark{1} Sakshi Gupta,\authormark{2} Bhagwati Prasad,\authormark{3} and Zubin Jacob,\authormark{1,2,*}}

\address{\authormark{1}Elmore Family School of Electrical and Computer Engineering, Purdue University, West Lafayette, 47907, IN, USA\\
\authormark{2}Department of Physics and Astronomy, Purdue University, West Lafayette, 47907, IN, USA\\
\authormark{3}Department of Materials Engineering, Indian Institute of Science, Bengaluru, 560012, Karnataka, India}

\email{\authormark{*}zjacob@purdue.edu} 


\begin{abstract*} 
The use of probabilistic spintronic devices for infrared radiation detection has introduced a shift in approach to thermal imaging. The integration of probabilistic magnetic tunnel junctions with infrared plasmonic nano-antennas yields digital-mode infrared sensors at room temperature.  Here, we present a scalable approach towards multi-pixel plasmonic-spintronic bolometer array fabrication and readout. We fabricate proof-of-concept 2x2 row-column multiplexed probabilistic plasmonic-spintronic arrays and show their response to long-wave infrared radiation (8-14 $\mu m$) with high readout speeds ($10^4-10^6$  counts per second). These spintronic, ultrafast, nanoscale (SUN) bolometers can enable high-pixel density CMOS compatible infrared detection platforms. Our work provides a broadband (9kHz to 3GHz) readout platform for future digital probabilistic detector applications. Furthermore, our approach addresses a key challenge associated with scaling infrared pixel sizes that can drive progress towards high pixel density detector arrays for infrared sensing and microscopy applications.

\end{abstract*}

\section{Introduction}
Historically, state-of-the-art long-wave infrared detectors have been dominated by HgCdTe photodiodes (cooled $\sim 77 $K) and VOx-based microbolometers (uncooled $\sim 300$K)\cite{rogalskiPixelScaling2016, rogalskinextdecade2017, EmergingIRTech2018}. These detectors have struggled to simultaneously achieve high sensitivity and high speed at room temperatures. In recent years, significant progress has been made in advancing infrared detector technology by incorporating
metastructures \cite{GoldBlack2016, liu2010infrared} to improve infrared absorption and exploring emerging material platforms such as graphene \cite{Graphene2018, graphene2022}, black phosphorus \cite{BlkPhosphorous2018}, etc. However, these material platforms have seldom unveiled multi-pixel array demonstrations for infrared applications. Recent demonstration of spin-based bolometry at room temperature using nanoscale spintronic devices \cite{SUNB} has
introduced a new paradigm of probabilistic detection for infrared imaging. This work
aims at extending this breakthrough digital-mode probabilistic technology to multi-pixel implementation
for thermal imaging and other emerging applications.

The recent demonstration of the Spintronic Ultrafast Nanoscale (SUN) bolometer has introduced the first digital-mode nano-bolometers at room temperature \cite{SUNB}. The detector uses the poisson response of stochastic Magnetic Tunnel Junctions (MTJs) to detect infrared radiation. This probabilistic nature of detection is different from conventional detection methods. The MTJ pillar is integrated with a plasmonic nano-antenna transduction layer to enhance infrared absorption by local surface plasmon resonances in the metasurface absorber and direct thermal energy towards the sensing structure \cite{zhou2021ultra,qin2021multi}. Recent advances in metamaterial absorber designs in the infrared also present opportunities for enhancing image contrast by suppressing stray light \cite{wu2026flexible} and towards polarization-sensitive infrared detection applications \cite{wu2026polarization}. 
The basic principle of detection uses thermally activated transitions between two stable magnetization states in stochastic MTJs to detect incident light. These transition events are called counts. The energy barrier between the two magnetization states can be manipulated by tuning the volume of the sensing layer. By designing the MTJ sensing layer's energy barrier ($E_b \sim 161 meV$) lower than the energy of a few infrared photons (248-1771 meV), the device can switch stochastically due to thermal heat from the surrounding. This rate of transitions can be modeled by the Neel-Arrhenius law \cite{Ohno_RTN_MTJ_2021}. Incident infrared radiation forms a hotspot, with heat propagating through the MTJ leading to an increase in the sensing layer temperature. This temperature increase results in a higher count rate demonstrating infrared sensitivity with a Noise-Equivalent Differential Temperature (NEDT) of 103mK \cite{SUNB}.  These nanoscale detector sizes (100nm to 300nm) can be used to implement oversampled and high-pixel density infrared arrays. Table \ref{tab:comparison} shows the comparison of SUN bolometers with existing state-of-the-art infrared detector technologies.

\begin{table*}[t]
\centering
\footnotesize
\renewcommand{\arraystretch}{1.2}
\setlength{\tabcolsep}{4pt}
\begin{tabular}{|>{\raggedright\arraybackslash}p{2.9cm}|>
{\raggedright\arraybackslash}p{2.7cm}|>{\raggedright\arraybackslash}p{2.5cm}|>{\raggedright\arraybackslash}p{3.4cm}|}
\hline
\textbf{Parameters} & \textbf{VOx Microbolometers} & \textbf{HgCdTe Photodiodes} & \textbf{SUN Bolometers} \\
\hline
\hline
\textbf{Thermal Sensitivity (NEDT) with frame rate} & 30--50 mK at 30 Hz \cite{rogalskiBook} & $<$\,20 mK at 30 Hz \cite{rogalskiBook} & 103 mK at 25 Hz \cite{SUNB} \\
\hline
\textbf{Operating Temperature} & 293 K & $\sim$77 K & 293 K \\
\hline
\textbf{Maximum Detector readout speeds} & 1-10 kHz \cite{abdel2017fabrication} & 1-400 MHz \cite{sun2017hgcdte,rothman2017hgcdte} & 0.1-458 Mcps \cite{SUNB} \\
\hline
\textbf{Typical Detector Size} & 8-50 $\mu$m \cite{rogalski2022scaling} & 5-20 $\mu$m \cite{rogalski2022scaling} & 100-300 nm device sizes \textit{(larger contact pads $\sim$ 25 $\mu$m, scalable below 1 $\mu$m \cite{lee20191gbit, alam2023versatile})} \\
\hline
\textbf{Electrical Power Consumption} & 0.1-10 $\mu$W \cite{chen2019microbolometric} & 0.1-10 nW \cite{tennant2014small} & 1-10 $\mu$W \cite{mousa2025neural} \\
\hline
\end{tabular}
\caption{Comparison of state-of-the-art infrared detectors with SUN bolometers.}
\label{tab:comparison}
\end{table*}

Over the past few decades, the number of pixels per infrared focal plane array (FPA) has been doubling every 18 months \cite{rogalskiPixelScaling2016}. Although array sizes will continue to grow, the rate of increase in pixel density is showing signs of slowing down\cite{rogalski2022scaling}. 
Scaling infrared pixels by reducing pixel sizes to obtain higher pixel density has been challenging. Pixel scaling in VOx-based microbolometers is limited by an increase in 1/f noise in the detector as the pixel area is reduced (pixel pitch $ \sim 8.5\mu m$) \cite{VOxPerformanceLimits2004,VOx_8um_pixel2023}. HgCdTe photodiodes have shown some scaling trends approaching sub-wavelength pixel sizes (pixel pitch $ \sim 5\mu m$) \cite{HgCdTeprogressTrends2015}. However, HgCdTe photodiodes are not CMOS back-end of line compatible which necessitates integration/hybridization with CMOS ROIC technology \cite{RogalskiHgCdTe2020Book}. Overall, the pixel response is also reduced as the detectors are scaled. Unlike existing detectors, spintronic bolometers \cite{SUNB} have recently been shown to operate with sub-wavelength pixel sizes (pixel pitch $< 1 \mu m$). As a result, spintronic nanobolometers can lead to interesting applications in near-field infrared sensing\cite{IR_near_field,MWIR_near_field} and infrared microscopy/microspectroscopy\cite{IR_microscopy_cancer,IR_microscopy_chemistry}.

In this paper, we demonstrate the implementation of a proof-of-concept row-column multiplexed array and readout of SUN bolometers. We fabricate 2x2 row-column multiplexed plasmonic-spintronic device arrays and show individual addressing of each pixel using a scalable broadband (9kHz to 3GHz) readout architecture. The high-speed nature of SUN bolometers is evident with device count rates over 50Kcps up to several Mcps.  We study and quantify the array's response to long-wave infrared (LWIR) radiation from a blackbody source by performing NEDT measurements on four devices. Thus, we extend the idea of spin-based bolometry to multi-pixel array implementations for high pixel density digital-mode imaging and other applications. This work reports the first demonstrated array implementation of row-column multiplexed stochastic MTJs with potential applications beyond infrared detection.

\section{Device Architecture and Operating Principle}
\begin{figure}[h!]
\includegraphics[scale=0.15]{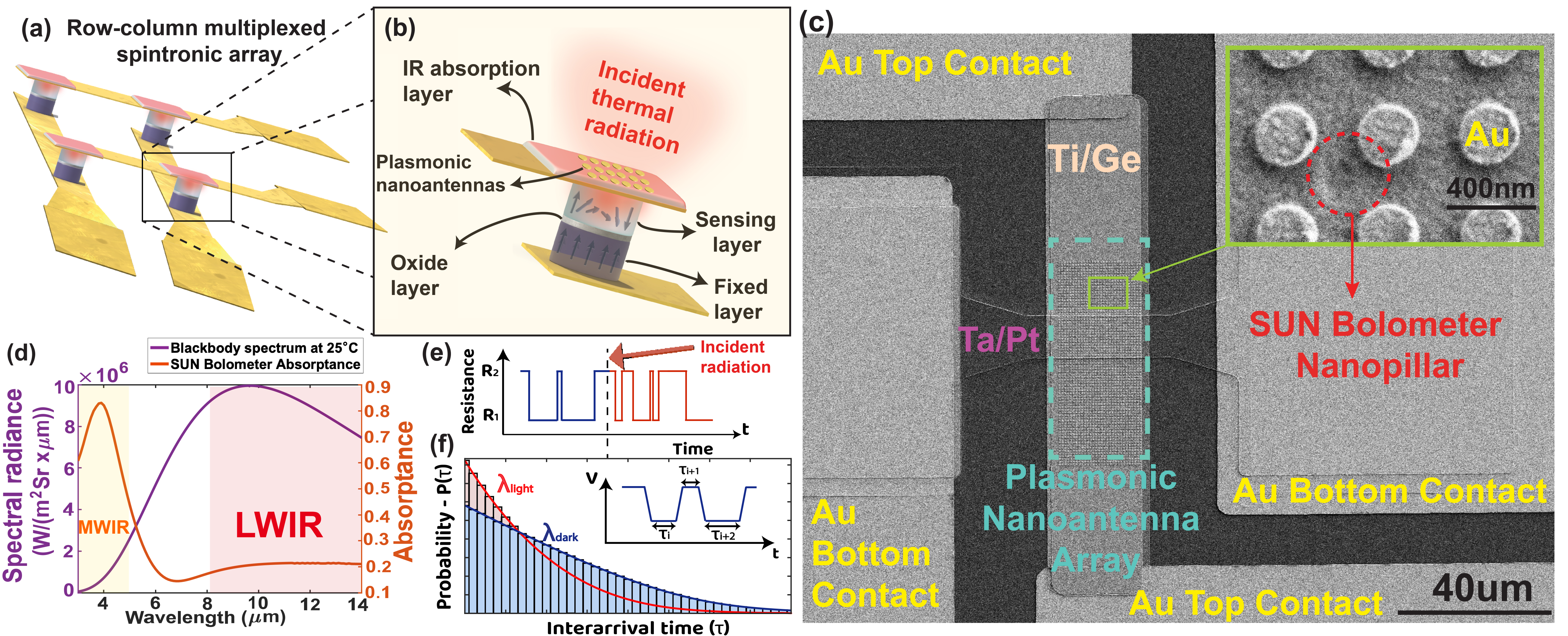}
\caption{\label{fig:Fig2_DevicePhysics} (a) The schematic of the row-column multiplexed spintronic bolometer array (b) SUN Bolometer device schematic demonstrating the operating principle of the device. The incident radiation is absorbed by the transduction layer. The transduction layer consists of plasmonic nanoantennas which enhances the absorbed radiation and directs heat towards the spintronic nanopillar. This heat causes an effective temperature increase in the sensing layer leading to higher rate of switching in the sensing layer. (c) Shows the SEM image of the single pixel SUN bolometer. (d) Measured absorptance of the SUN bolometer plasmonic nanoantenna array (orange) compared to 300K black body spectral radiance (purple). (e) The magnetization flips of the sensing layer are readout as transitions between two resistance states of the device. These transitions increase as the temperature in the sensing layer increases. (f) Histogram of the count interarrival times (i.e. the time between two nearest count events). The curve exhibits an exponential dependence indicating that the count statistics follow a poisson process. Inset shows the measurement of interarrival times for a readout signal.
}
\end{figure}

Fig. \ref{fig:Fig2_DevicePhysics}(a) shows a schematic of the proposed row-column multiplexed spintronic array for longwave infrared bolometry. Fig. \ref{fig:Fig2_DevicePhysics}(b) depicts the schematic of a single pixel SUN bolometer device consisting of the stochastic MTJ device with a transduction layer incorporating gold plasmonic nanoantennas on germanium. The SEM image of the SUN bolometer device is shown in Fig. \ref{fig:Fig2_DevicePhysics}(c) where the Ti/Ge transduction layer with the Au nanoantenna array can be seen. The inset shows the location of the MTJ nanopillar of the SUN bolometer. The absorption measurements of the transduction layer  displays significant infrared absorption as shown in Fig. \ref{fig:Fig2_DevicePhysics}(d). The heat transfer from the transduction layer to the MTJ nanopillar results in an increase in transitions between parallel and anti-parallel magnetization states of the device. This can result in sub-micron active area infrared devices opening doors for sub-wavelength infrared detection.

The parallel and anti-parallel magnetization states manifest as two stable resistance states on the device readout owing to the tunnel magnetoresistance of the MTJ as depicted in Fig. \ref{fig:Fig2_DevicePhysics}(e). These transitions in the device resistance can be processed to get the device count rate (or change in the count rate). The inherently digital-mode process of bistable magnetization flips follow poisson statistics which is missing in analog detection mechanisms. The histogram statistics of the interarrival times (time between two counts) is an exponential indicating the events follow a poisson process. Fig. \ref{fig:Fig2_DevicePhysics}(f) shows the simulated change in the equilibrium distribution due to the change in the count rate induced by incident infrared radiation on the device.

\section{Array Fabrication and Readout Architecture}

\begin{figure}[h!]
\includegraphics[scale=0.31]{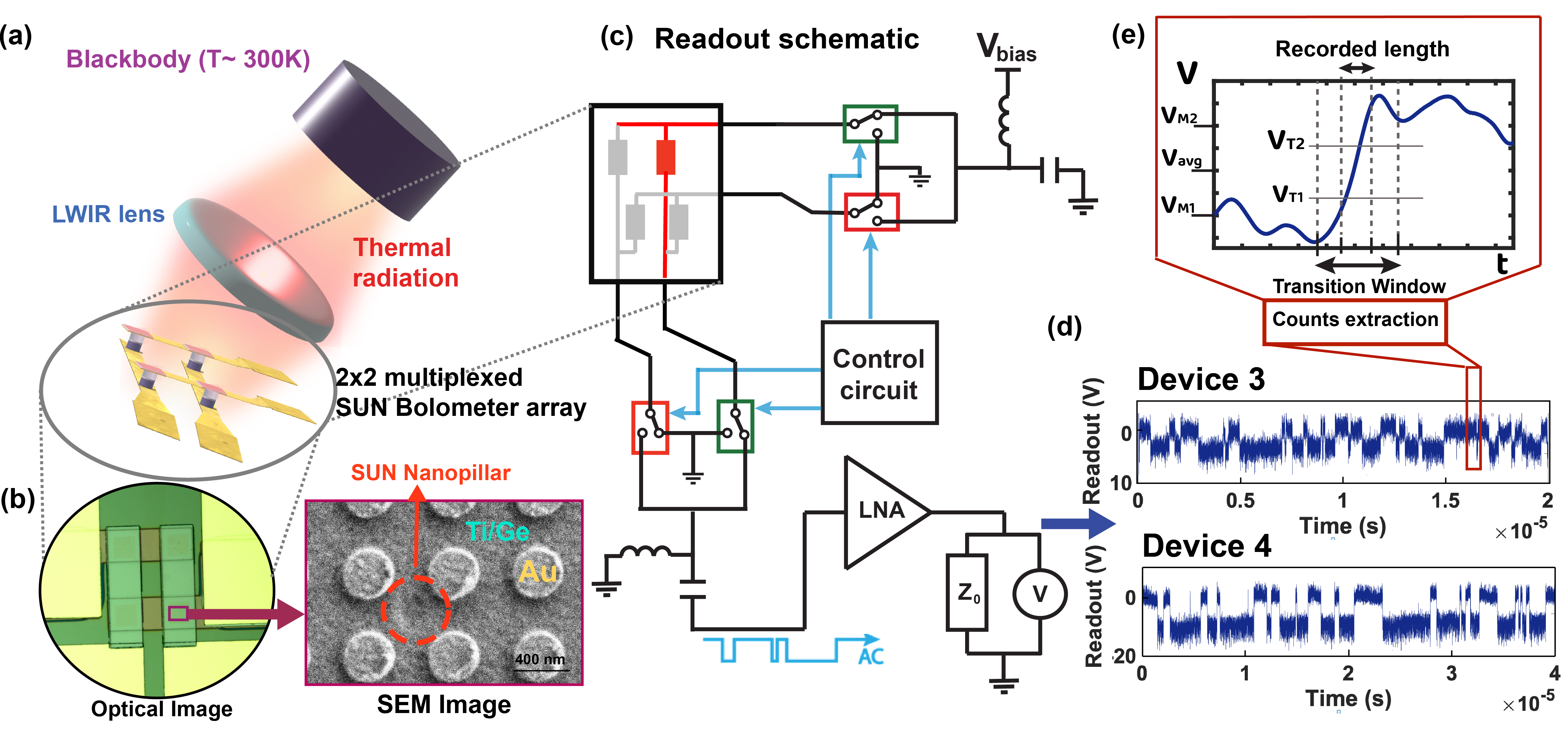}
\caption{\label{fig:Fig3_Readout}(a) Simplified schematic of row-column multiplexed array in a basic imaging setup with a blackbody source. (b) Optical image of a 2x2 SUN bolometer array and SEM image of a single SUN bolometer with plasmonic integration (c) Readout schematic of the 2x2 multiplexed array- To select a device, the corresponding row and column are addressed using RF switches controlled by a digital control circuit. All the other devices are connected to AC ground on both terminals reducing leakage or splitting of their signal in the array. Through the RF switches, one row is biased with an input voltage and one of the column lines is connected to the readout path. The row line is AC grounded and the column line is DC grounded using a bias-tee. This ensures that the AC signal from only one device is readout by the column readout path. The other devices are either unbiased or have AC ground connected to both terminals. (d) Experimental readout waveform of two devices on the same array.
(e) Count extraction algorithm where suitable thresholds and time windows are set to capture these ultrafast transition events.
}
\end{figure}

To implement spintronic multipixel arrays, we fabricate a row-column multiplexed proof-of-concept SUN bolometer array with a scalable readout platform. We fabricate nanoscale 2x2 SUN bolometer device arrays (Supplementary section I). The devices are multiplexed so that one terminal of each device is connected to the row line and the other is connected to the column line as shown in the schematic and optical image in Figs. \ref{fig:Fig3_Readout}(a,b). 
We also show individual addressing of each pixel using a dedicated readout architecture presented in Fig. \ref{fig:Fig3_Readout}(c). We connect each row/column line to either the readout path or the ground using broadband RF switches. Additionally, using bias-tees we create AC and DC grounds on the row and column lines respectively. This arrangement grounds the signal from other pixels and allows the readout of one pixel through the readout path as shown in Fig. \ref{fig:Fig3_Readout}(c). It is important to have low signal leakage from devices in the array into readout of other devices. In our readout platform, signal leakage due to other devices (crosstalk) is reduced by the interplay of AC and DC grounds in the row and column lines. As shown in Fig. \ref{fig:Fig3_Readout}(c), the devices that are not being read out are AC grounded on both terminals minimizing their leakage into the main readout path.

This row–column multiplexed platform enables tighter pixel packing, allowing sub-micrometer pixel pitch for infrared detectors. The readout circuit can also provide high amplification and well capacity due to larger available area outside the array. Although sequential readout is demonstrated, the architecture can support simultaneous row-wise readout by replicating column readout paths for each column. We have included quantitative analysis of this crosstalk as the array size is increased (see supplementary section IV). We show that this architecture is scalable to kilo-pixel arrays and beyond when it comes to signal leaking from other devices.

Fig. \ref{fig:Fig3_Readout}(d) shows the experimental readout response of two stochastic devices belonging to the same 2x2 multiplexed array. However, due to low-yield we demonstrate response of four working devices from three different 2x2 arrays. To get the device count response, we employ a thresholding algorithm to capture these ultrafast transitions and integrate them together to get the count rates of the devices. A visualization of this thresholding algorithm is shown in Fig. \ref{fig:Fig3_Readout}(e). 
We run this threshold check over a window and find the fastest transition between the two levels in that window. We characterize the stochastic response of these devices by establishing a magnetic field dependence (Supplementary section I). This concludes that the device response is  magnetoresistive. As it is evident from the magnetic field-dependent nature of the device response, an external magnetic field bias is required to optimally bias the device to get a high-count response. This magnetic field is applied to compensate for the inherent stray field in the sensing layer of the device. Therefore, the externally applied magnetic field can directly tune the device response to incident radiation (for more details, see supplementary information section s1-B).

\section{Response to Long-Wave Infrared Radiation}
To study and quantify the array's response to LWIR radiation, we perform NEDT measurements on four devices. NEDT is the minimum temperature change that the sensor can detect. While Noise-Equivalent Power (NEP) is a widely used metric to characterize detectors, it does not perform equally well for thermal measurements where we attempt to distinguish changes in temperature of a black body source from some background. This can be attributed to NEP measurements being performed at a single wavelength and only account for the inherent detector noise (dark noise). On the other hand, NEDT is a broadband measurement and accounts for source noise (shot noise), temperature fluctuations, background infrared noise in addition to the detector noise. Therefore, NEDT is a key performance metric for thermal imaging sensitivity, frequently employed in commercial imagers. It is given by: \begin{equation}
 NEDT =  \frac{C_n}{(dC/dT)}, 
 \end{equation} 
where $C_n$ is the detector noise given by standard deviation of the detector count response at a particular temperature. dC/dT is the change in detector counts measured with respect to the change in blackbody source temperature. For digital-mode detectors the signal is given by the mean counts across several measurements. The detector noise is calculated by taking the standard deviation between multiple measurements. 


\begin{figure}[h!]
\centering
\includegraphics[scale=0.37]{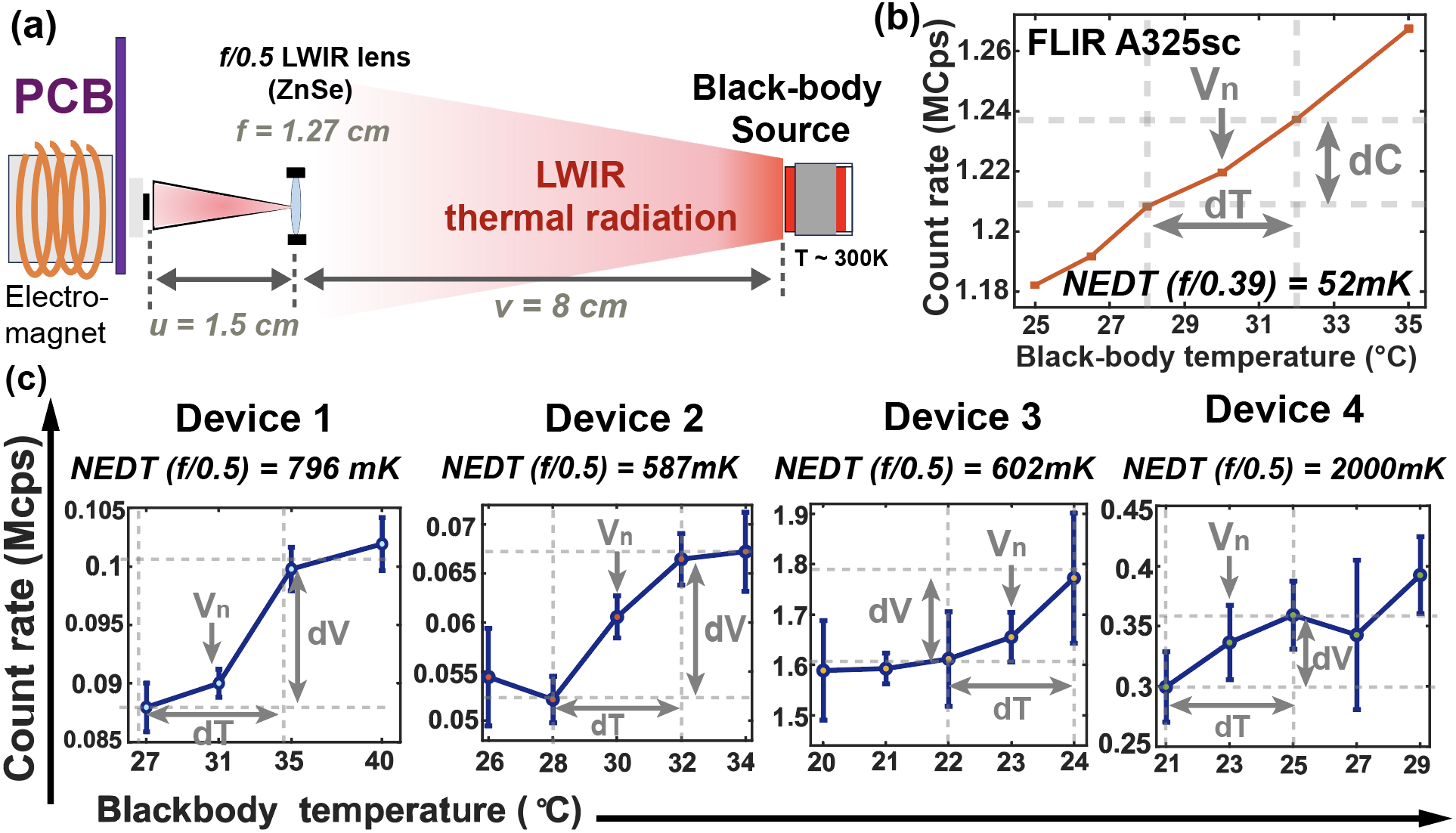}
\caption{\label{fig:Fig4_NEDT}(a) NEDT measurement setup with f/0.5 LWIR optics. The extended array blackbody source has a stability of 1 mK. To measure NEDT the blackbody source is set to a fixed temperature and the device response is measured using 5 sequential measurements. Then the blackbody temperature is changed and device is measured again. (see Supplementary section I, subsection C) (b) The setup is characterized using FLIR A325sc thermal camera. This characterization benchmarks and validates our measurement setup using an existing infrared detector. (c) Response of 4 SUN bolometer array devices to thermal radiation in NEDT measurements. The linear response and standard deviation are used to find NEDT. SUN bolometer error bars represent standard deviation of 5 measurements with 20ms integration time (50 Hz). The four devices are biased with the following voltage and magnetic field biases: Device 1(-230 mV, 125 Oe), Device 2(234 mV, -865 Oe), Device 3(-181 mV, 15 Oe), Device 4(180  mV, 2 Oe).
}
\end{figure}

Fig. \ref{fig:Fig4_NEDT}(a) shows the standard experimental setup used to measure NEDT. We characterize the setup using a commercial VOx-based microbolometer imager- FLIR A325sc by benchmarking the NEDT measurement shown in Fig. \ref{fig:Fig4_NEDT}(b). This measurement validates our NEDT setup with existing state-of-the-art uncooled detector technology. Using this setup, we measure the thermal response of our devices using f/0.5 aperture LWIR lens at room temperature. Fig. \ref{fig:Fig4_NEDT}(c) shows the response of the four SUN bolometer devices. We have discussed our NEDT measurement setup, measurement details and limitations in supplementary section s1-C. We have also discussed the analysis of repeated measurements, uncertainty ranges and dependence of NEDT with integration time in Supplementary Information (see section s1-C).  Due to saturation in our devices, the device response has a lower linearity compared to the highly linear response of VOx-based microbolometer in the FLIR camera.  Each device has a different response and NEDT due to variations in the fabrication process pertaining to material deposition, annealing, transduction layer integration/alignment and controlling out-of-plane micro-domain formations of in-plane sensing layer in devices with out-of-plane readout layer (see Supplementary section I, subsection D). With improvements in fabrication and using different thin-films having in-plane sensing and in-plane readout layers, we can get more uniform response from these devices. Moreover, obtaining higher count rate devices can result in better NEDT around 103mK at 25Hz as recently demonstrated for a single-pixel device \cite{SUNB}.

\section{Discussion}

\begin{figure}[b!]
\centering
\includegraphics[scale=0.4]{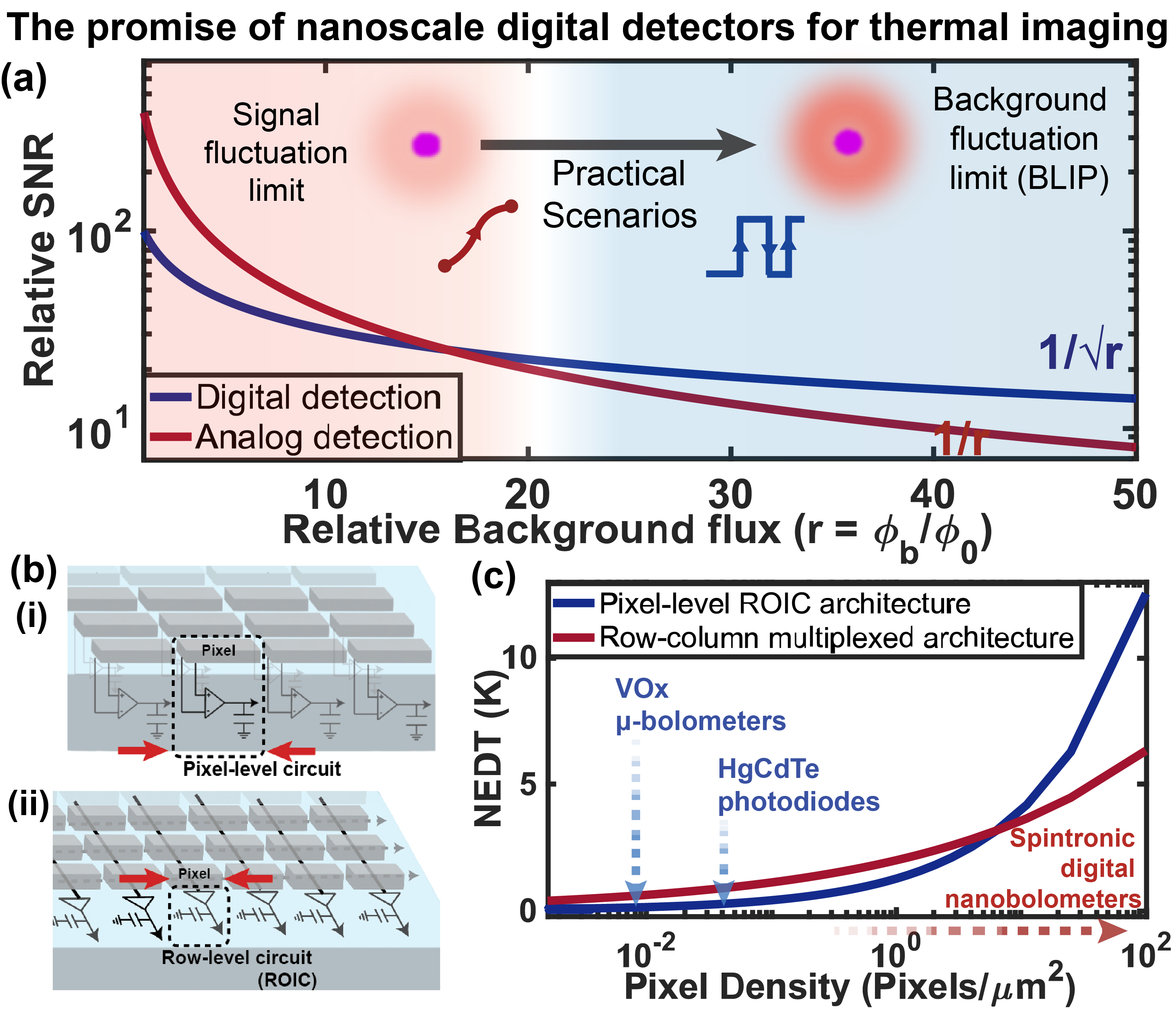}
\caption{\label{fig:Fig1_DigitalvAnalog} (a) Advantage of digital detection mechanism for capturing the signal in high background scenarios (Supplementary section I). (b) Comparison of readout architectures for infrared arrays - (i) Schematic representation of pixel level ROIC architecture. (ii) Schematic of proposed row-column multiplexed architecture. (c) Comparison of the NEDT trade-offs with the two architectures (supplementary sII). The dashed arrows mark the smallest detector devices currently fabricated in comparison with spintronic nanobolometers.
}
\end{figure} 
As discussed and demonstrated, the spintronic devices we fabricate and characterize are sub-micron in size. Our approach towards developing smaller sub-wavelength pixels (below $8 \mu m$) for infrared imaging can have its own set of advantages beyond aliasing and resolution. This proposed approach can lead to improvements in sensitivity and reduction of false alarm rates using pixel correlations \cite{SmallPixelOversampled}. Previous analyses of detectors for applications in astronomical FPAs have shown that reducing pixel sizes lowers the Cramér–Rao bound for astrometric precision. ($\sigma_{CR}^2 \sim pitch$) \cite{SmallPixel_background_astronomy}. Moreover, in the small-pixel regime, background radiation can significantly impact the accuracy of detection increasing the cramer-rao bound ($\sigma_{CR_{oversampled}}^2 \sim Background$)\cite{SmallPixel_background_astronomy}. \\
In the context of infrared background, we now discuss the advantage of digital mechanism of the SUN bolometer over its analog counterparts. Infrared detectors are often background-limited and operate in the background fluctuation limit or the Background-limited infrared photodetector (BLIP) regime \cite{rogalskiOpticalDetection2017}. This limit is achieved when the background noise exceeds the device noise. For detectors operating at room temperature, the background is significantly higher compared to cooled detectors \cite{yang2011investigation}. In Fig. \ref{fig:Fig1_DigitalvAnalog}(a), we show an improved signal-to-noise ratio (SNR) trend of digital detectors over analog detectors for high background scenarios. It is established that the digital mechanism of detection shows a poisson response to incident light \cite{SUNB,SPADPoisson, PoissonDigital}. The SNR for digital detectors ($SNR \propto 1/\sqrt{\phi_B}$) drops less severely in comparison to analog detectors ($SNR \propto 1/\phi_B$) as background increases (Supplementary section II).  Furthermore, digital detectors have demonstrated opportunities in low-power sensing \cite{LowPowerIR_MEMS2017} and background reduction using pixel correlations\cite{SPADBackground2018}.


Scaling infrared pixel sizes requires shrinking both the detector and its associated circuit. Although reducing the detector size is difficult, the readout mechanism often becomes the main bottleneck. Modern FPAs use pixel-level readout, where each pixel has its own circuit to integrate the electrical signal. A representative schematic is presented in Fig. \ref{fig:Fig1_DigitalvAnalog}(b-i). This circuit usually integrates the signal over a capacitive memory element, and the maximum integration capacity can impact the sensitivity of the pixel. This is described in terms of the well capacity ($N_w$) of the detector. As the pixel area scales, this well capacity reduces ($N_w \sim A_{pixel}$), impacting the overall pixel sensitivity (Supplementary section III). Therefore, for applications involving very high pixel density, the overall readout architecture has to be re-evaluated.\\
Our proposed row-column multiplexed approach for infrared detector arrays incorporates a high density of devices in a small area as shown in Fig. \ref{fig:Fig1_DigitalvAnalog}(b-ii). This can result in scaling infrared detectors which is ultimately only limited by the detector and interconnect area. Due to shared circuitry by each row, this approach permits the readout of one row at a time reducing the integration time per row by 1/n (for $n \times n$ array). However, it also results in significantly higher amplification and well capacities because the readout circuit is not restricted by the pixel area anymore. Fig. \ref{fig:Fig1_DigitalvAnalog}(c) highlights the trend that, beyond a certain pixel density, the compromise on sensitivity due to the pixel-level ROIC architecture would be significant compared to the row-column multiplexed architecture (Supplementary section III). This provides a platform for ultra-high pixel densities suitable for near-field IR and microscopy applications. Through this discussion, we present some important merits towards having spintronic digital detectors that can be scaled to ultra-high pixel densities (sub-$\mu m^2$ pixels).
\\

In conclusion, we show response of the multi-pixel probabilistic spintronic bolometer array to LWIR radiation. We demonstrated a row-column multiplexed array that can be used to achieve high pixel density using nanoscale bolometers. Furthermore, this also serves as a readout platform for future stochastic bolometers and stochastic computing devices \cite{mousa2025neural}. Our demonstration and analysis of the multiplexed arrays and readout platform implies that stochastic spintronic arrays show promise to be used for high-speed and high-resolution infrared detection applications. Consequently, this can also result in progress towards near-field and microscopy applications across the IR spectrum \cite{mousa2026ultra}.

\begin{backmatter}

\bmsection{Acknowledgment}
We would like to thank Dr. Tiffany Santos from Western Digital for providing the thin-films used to fabricate the devices in this work. This work was partially supported by an Elmore Chaired Professorship at Purdue University. 

\bmsection{Disclosures}
The authors declare no conflicts of interest.

\bmsection{Data Availability Statement}
Data underlying the results presented in this paper are not publicly available at this time but may be obtained from the authors upon reasonable request.

\bmsection{Supplemental document}
For more information about this work, please refer to supplementary information.

\end{backmatter}
\bibliography{references}








\end{document}